\documentclass[usenatbib]{mn2e}
\usepackage{graphicx}

\def\apj{\rm ApJ}
\def\apjl{\rm ApJL}
\def\apjs{\rm ApJS}
\def\aj{\rm AJ}
\def\mnras{\rm MNRAS}
\def\nat{\rm Nature}
\def\pasj{\rm PASJ}
\def\pasp{\rm PASP}

\def\aap{\rm AAP}
\def\araa{\rm ARA\&A}

\def\gax{\mathrel{\raise.3ex\hbox{$>$}\mkern-14mu\lower0.6ex\hbox{$\sim$}}}
\def\lax{\mathrel{\raise.3ex\hbox{$<$}\mkern-14mu\lower0.6ex\hbox{$\sim$}}}
\def\gtorder{\mathrel{\raise.3ex\hbox{$>$}\mkern-14mu
             \lower0.6ex\hbox{$\sim$}}}
\def\ltorder{\mathrel{\raise.3ex\hbox{$<$}\mkern-14mu
             \lower0.6ex\hbox{$\sim$}}}

\begin{document}

\title [Dust and Supernova Binary Companions]
   {Dust Formation and the Binary Companions of Supernovae}

\author[C.~S. Kochanek]{ 
    C.~S. Kochanek$^{1,2}$, 
    \\
  $^{1}$ Department of Astronomy, The Ohio State University, 140 West 18th Avenue, Columbus OH 43210 \\
  $^{2}$ Center for Cosmology and AstroParticle Physics, The Ohio State University,
    191 W. Woodruff Avenue, Columbus OH 43210 \\
   }

\maketitle

\begin{abstract}
Supernovae (SNe) should both frequently have a binary companion at death 
and form significant amounts of dust.  This implies that any binary
companion must lie at the center of an expanding dust cloud and
the variable obscuration of the companion as the SN remnant 
(SNR) expands will both unambiguously mark the companion and allow the 
measurement of the dust content through absorption rather than emission for decades
after the explosion.  However, sufficiently hot and luminous companions can
suppress dust formation by rapidly photo-ionizing the 
condensible species in the ejecta.  This provides a means of
reconciling the Type~IIb SNe Cas~A, which lacks a luminous companion and
formed a significant amount of dust ($M_d \gtorder 0.1 M_\odot$), with
the Type~IIb SNe~1993J and 2011dh, both of which appear to have 
a luminous companion and to have formed a negligible amount of dust 
($M_d \ltorder 10^{-3}M_\odot$). The Crab and SN~1987A are consistent
with this picture, as both lack a luminous companion and formed
significant amounts of dust.  An 
unrecognized dependence of dust formation on the properties of binary companions
may help to explain why the evidence for dust formation in SNe appears
so contradictory.
\end{abstract}


\section{Introduction}
\label{sec:introduction}

The role of binaries and dust formation are two of the perennial questions and challenges
for understanding the properties of supernovae (SNe) and their consequences.  Binaries
are common (e.g., \citealt{Sana2012}, \citealt{Duchene2013}, \citealt{Kobulnicky2014},
\citealt{Moe2016}) and can modify stellar evolution,
likely invalidating many expectations for the properties of SN and their progenitors
based on the evolution of isolated stars (e.g., \citealt{Eldridge2008}, \citealt{Sana2012}).
Their role would be much better understood
if it were possible to cleanly survey SNe, their progenitors, or their remnants for 
binary companions and to then well-characterize the overall population.  SNe are also
believed to be an important source of dust, along with asymptotic giant branch (AGB)
stars, particularly early in the universe where there is little time for stars to
evolve to the AGB (e.g., \citealt{Gall2011}, \citealt{Cherchneff2014}).  However, where dust 
formation in AGB stars is relatively easy to
measure, both the amount of dust formed in SNe and the fraction of that dust which
survives to be mixed into the interstellar medium  remain open questions.   

Searches for stellar companions to SN have largely focused on Type~Ia SN
as a means of distinguishing the single and double degenerate models (see
the review by \citealt{Maoz2014}).
The original picture was relatively simple: in the single degenerate model 
there should always be a companion and in the double generate model there 
should never be a companion.  The prevalence of triple systems and the potential role of
Kozai-Lidov oscillations in producing double degenerate Type~Ia SN
complicates the latter
case since it means that double generate Ia's may be genuinely associated
with non-degenerate (tertiary) companions (e.g., \citealt{Thompson2011}, \citealt{Kushnir2013}).  

Searches for companions can either be direct, simply observing
the companion, or indirect, observing other consequences of the companion's
existence.  Direct searches for companion stars to Type Ia SNe include examining the pre-explosion 
data for SN~2011fe (\citealt{Li2011}, \citealt{Kelly2014}) and looking in 
supernova remnants (SNR) such as Tycho (\citealt{Ruiz2004}, \citealt{Ihara2007}),
SN~1006 (\citealt{Schweizer1980}, \citealt{Gonzalez2012}) and 
SNR~0509--67.5 (\citealt{Schaefer2012}).  Indirect searches
include the effects of close companions on early-time SN
light curves (e.g., \citealt{Kasen2010}), searches for narrow
hydrogen emission lines from material stripped from the
companion (e.g., \citealt{Leonard2007}, \citealt{Shappee2013}), and searches for 
radio emission as the SN shock passes through the wind 
from a non-degenerate companion (e.g., \citealt{Chomiuk2016}).

Far less observational attention has been given to the binary companions to core collapse
SN (ccSN), although \cite{Kochanek2009} estimated that 50-80\% of ccSN are
probably members of a stellar binary at death.  More generally, most massive
stars are in binaries and many ccSN progenitors should be the
remnants of stellar mergers, have undergone mass transfer or simply have a binary
companion (e.g., \citealt{Sana2012}, \citealt{Kobulnicky2014}, \citealt{Moe2016}).  
For example, the numbers of stripped Type~Ibc SNe and the limits on
their progenitor stars both suggest that many are stripped through binary mass
transfer rather than simply wind (or other) mass loss (e.g., \citealt{Eldridge2008},
\citealt{Smith2011}, \citealt{Eldridge2013}). Most theoretical studies 
(e.g., \citealt{Yoon2010}, \citealt{Yoon2012}, \citealt{Yoon2017},
\citealt{Claeys2011}, \citealt{Benvenuto2013}, 
and \citealt{Kim2015}) have focused on the binary properties of the stripped ccSN classes 
(Type~IIb, Ib and Ic) as cases where binary evolution is modifying the outcomes.
As with the Type Ia SN, companions can modify the early-time light curves of
the SN (e.g.,  \citealt{Kasen2010}, \citealt{Moriya2015}, \citealt{Liu2015}). 
Searching for narrow hydrogen emission lines in the nebular phase is less 
promising even for Type~Ib/c SNe because of the short stellar life times and the role of winds in
stellar mass loss. Because luminosities increase so rapidly with mass, 
the signatures of shock heated companions to ccSNe will also be weaker than
those for companions to Type Ia SNe.   

There is no certain identification of a binary companion to a ccSNe. The Crab,
Cas~A, and SN~1987A lack luminous companions, with upper mass limits of order
$1$-$2M_\odot$ (\citealt{Kochanek2017}).  \cite{Graves2005} found somewhat 
tighter limits for SN~1987A, but assumed far less dust than is now known
to be present (see below).  The Crab was a low-energy Type~II
SN, possibly an electron capture SN (e.g., \citealt{Hester2008},
\citealt{Smith2013}), Cas~A was a Type~IIb SN (\citealt{Krause2008},
\citealt{Rest2008}), and SN~1987A was a (peculiar?) Type~II SN (\citealt{Arnett1989}).     
The best case for a detection is probably a hot, luminous companion
($T\sim 20000~$K, $L\simeq 10^5 L_\odot$) to the Type~IIb SN~1993J 
(\citealt{Maund2004}, \citealt{Fox2014}).  The existence of a similar
companion to the Type~IIb SN~2011dh is debated, with \cite{Folatelli2014} arguing
for a detection and \cite{Maund2015} arguing that the flux may be 
dominated by late time emission from the SN.  There is some
evidence of a blue companion for the Type~IIb SNe~2001ig (\citealt{Ryder2006})
and SN~2008ax (\cite{Crockett2008}.
There are limits on the existence companions to the Type Ic SNe~1994I (\citealt{VanDyk2016})
and SN~2002ap  (\citealt{Crockett2007}) and the Type~IIP 
SN~2005cs (\citealt{Maund2005}, \citealt{Li2006}),
and SN~2008bk (\citealt{Mattila2008}).  All the claimed detections are blue, 
which is expected for most binary companions (\citealt{Kochanek2009}). However,
in the presence of dust formation, early-time limits on the existence of binary 
companions are problematic, as we discuss below.

For either SN class, the challenge is to prove that any candidate is a
real binary companion with the further complication of triples, particularly
for the single versus double degenerate debate.  Pre-supernova light curves
can identify close binaries, but data of adequate depth and cadence
are only beginning to exist (e.g., \citealt{Kochanek2012},
\citealt{Kochanek2016}) and any detections depend on rare,
favorable geometries.  Stars identified after the SN may
be distinguishable due to being shock heated and hence
over-luminous (e.g., \citealt{Marietta2000},
\citealt{Shappee2013b}, \citealt{Pan2014}). Other signatures
may be peculiar kinematics (\citealt{Ruiz2004}), 
chemical abundances (e.g., \citealt{Ihara2007}, \citealt{Gonzalez2009},
\citealt{Kerzendorf2009}, \citealt{Kerzendorf2013}) or fast
rotation rates (e.g., \citealt{Kerzendorf2009}, \citealt{Pan2012}, \citealt{Kerzendorf2013},
\citealt{Pan2014}).  Finally, the supernova ejecta can produce broad
absorption lines in the spectra of stars either inside or behind the SNR (e.g., 
\cite{Wu1993} for the star behind SN~1006 identified by \cite{Schweizer1980}
and the remnant of SN~1885 in M31 found by \cite{Fesen1989}).
All these spectroscopic tests can also
be applied to the companions of ccSN even though
they have been discussed almost exclusively in the context of Type~Ia SN.
All spectroscopic tests are, however, challenging to apply in an 
extragalactic context because they require high resolution, high signal-to-noise 
spectra of the candidate star.  Ideally, we need a time variable signature for
a binary companion that can be carried out with broad band photometry for
decades after the SN.

Dust formation is predicted for most SN, with the total amount produced
diminishing with decreasing ejecta masses and increasing explosion 
energies because these lead to higher velocities and more rapidly
dropping densities (e.g., \citealt{Dwek1988}, \citealt{Kozasa1989}
\citealt{Todini2001}, \citealt{Cherchneff2010}, \citealt{Nozawa2010},
\citealt{Nozawa2011}, \citealt{Sarangi2015}).  Typical theoretical estimates 
are that ccSN produce $M_d \sim 0.1$-$1 M_\odot$ of dust while
Type~Ia produce $M_d \sim 10^{-3}$-$0.1 M_\odot$ of dust.  These predictions
appear to be consistent with late time observations of SNR but such
high dust masses are rarely seen shortly after a SN (see the 
reviews by \cite{Gall2011} or \cite{Cherchneff2014}).
Ignoring SN with strong circumstellar interactions, evidence for
dust in SN is usually found roughly a year or two after the explosion
as either a blue shift of the emission lines or a mid-IR excess
due to the presence of hot dust.  Examples include SN~1987A 
(\citealt{Wooden1993}), SN~2003gd (\citealt{Sugerman2006}),
and SN~2004et (\citealt{Kotak2009}).  

Observations of SNR
at much later times (decades or longer), particularly at
longer wavelengths corresponding to emission by cooler dust,
generally find dust masses more compatible with theoretical
predictions.  Recent examples include SN~1987A (\citealt{Matsuura2011},
\citealt{Indebetouw2014}, \citealt{Matsuura2015}), Cas~A (\citealt{Barlow2010},
\citealt{deLooze2016}) and the Crab (\citealt{Temim2013},
\citealt{Owen2015}).  These higher dust masses may not be
universal, as \cite{Temim2010} find only a few $10^{-2} M_\odot$ 
of dust in the pulsar wind-containing remnant G54.1$+$0.3.
There is little evidence for cold dust in the Tycho or Kepler
Type~Ia remnants ($M_d < 10^{-2}M_\odot$, \citealt{Gomez2012}).
A fundamental problem for all these estimates is that they depend on identifying
dust by emission.  Observations during the SN are generally only 
sensitive to hot dust because they depend on observations at relatively
short mid-IR wavelengths (e.g., the Spitzer $3.6$ and $4.5\mu$m bands), while
searches for cold dust require observations at long wavelengths where
the combination of luminosities and
sensitivities usually restrict the searches to the Local Group
or just to the Galaxy and Magellanic Clouds.  

In theory, these two problems can solve themselves -- the time variable 
absorption from the dust produced by a SN is a nearly unique signature 
of a binary companion, and the absorption of light from the companion 
by the dust is a 
temperature-independent probe of the amount of dust formed.   
The observations are 
also feasible for decades after the SN, permitting relatively
large surveys for both dust and binary companions at the sites
of SN limited only by the luminosity of the companions.  
In \S2 we introduce a simple model for the dust and discuss
its implications for specific SNe.  There is, however, a caveat.
As we discuss in \S3, a sufficiently hot and luminous binary
companion can suppress dust formation, which may explain the
difference between the Type~IIb SNe Cas~A (lots of dust and no
luminous companion) and SN~1993J (no dust and a luminous
companion).  In \S4 we use the simple
model for binary companions to SNe from \cite{Kochanek2009}
to illustrate the effects of dust on the detectability of
binary companions.  We end in \S5 with a general discussion
and potential observations.

\begin{figure}
\centering
\includegraphics[width=0.45\textwidth]{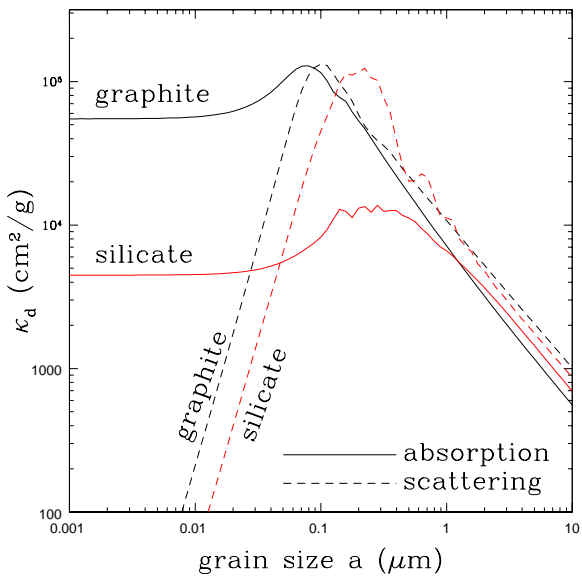}
\caption{ Absorption ($\kappa_a$, solid) and scattering ($\kappa_s$, dashed) opacities for \protect\cite{Draine1984} 
  graphitic (black) and \protect\cite{Laor1993} silicate (red) dust as a function of size $a$. We scale
  the results to $\kappa = 10^4 \kappa_4$~cm$^2$/g.
  }
\label{fig:opacity}
\end{figure}

\begin{figure}
\centering
\includegraphics[width=0.45\textwidth]{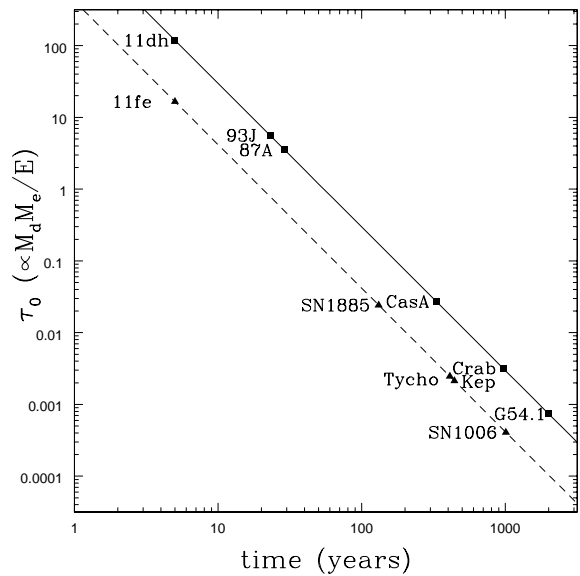}
\caption{ Expected visual optical depths ($\tau_0$, Equation~\protect\ref{eqn:tau0})  as a function of time for ccSNe
  (solid) and Type~Ia (dashed) SN. The rough expectations for the sources discussed in the text
  are indicated by squares and triangles for ccSNe and Type~Ia SNe, respectively. The
  scalings assume a dust mass of $M_d=0.1M_\odot$, an explosion energy of $E=10^{51}$~erg,
  a visual opacity of $\kappa=10^4$~cm$^2$/g (see Figure~\protect\ref{fig:opacity}) and
  ejecta masses of either $10M_\odot$ (ccSNe) or $1.4M_\odot$ (Type~Ia).
  }
\label{fig:expected}
\end{figure}

\section{Models}

We consider a simple self-similar (e.g., \citealt{Chevalier1982}) 
model for the SN ejecta density distribution,
\begin{equation}
   \rho(r,t) = { 15 M_e \over 32 \pi v_0^3 t^3 } \left( { r \over v_0 t } \right)^{-x}
  \label{eqn:selfsim}
\end{equation}
where $ x=0$ ($x=8$) for $v < v_0$ ($v>v_0$) or, equivalently, $r < v_0 t$ ($r > v_0 t$).  
The total kinetic energy of the debris of
$E = M_e v_0^2 /2$ is determined by the ejecta mass $M_e$ and the velocity scale
$v_0$.  By mass, $5/8$ of the ejecta lies in the interior region and $3/8$ lies 
in the exterior region.  For an energy scale of $E= 10^{51}E_{51}$~erg and
a mass of $10 M_\odot$ ($1.4 M_\odot$), the characteristic velocity is 
$v_0 \simeq 3200$~km/s ($v_0 \simeq 8500$~km/s) for a Type~IIP (Type~Ia) SN. 
We adopt these simple models with $E_{51}=1$ for our standard results.   
The break radius scale is
\begin{equation}
    r_0 = v_0 t = 9.5 t_{100} v_{3000} \times 10^{17}~\hbox{cm} 
    \label{eqn:size}
\end{equation}
where $t = 100 t_{100}$~years and $v= 3000 v_{3000}$~km/s.  The size of the
ejecta cloud grows sufficiently rapidly that we can simply view any binary
companion as lying at the center of the dust distribution.

If the ejecta contains dust with a total mass of $M_d=0.1 M_{d0.1}M_\odot$ 
which is uniformly mixed with the gas and has
a constant opacity of $\kappa = 10^4 \kappa_4$~cm$^{2}$/g, then 
the optical depth scale is
\begin{equation} 
   \tau_0 = { 15 \kappa M_d \over 32 \pi t^2 v_0^2 }  
          = 0.33 \kappa_4 M_{d0.1} v_{3000}^{-2}  t_{100}^{-2}. 
     \label{eqn:tau0}
\end{equation}
Alternatively, we can replace
the ejecta velocity with the kinetic energy to find that
\begin{equation} 
   \tau_0 = { 15 \kappa M_d M_e \over 64 \pi E t^2 }  
          = 0.30 \kappa_4 M_{d0.1} M_{e10} E_{51}^{-1} t_{100}^{-2}.
\end{equation}
For a fixed dust mass, Type~Ia SN will have less optical depth
at any given time due to their significantly faster expansion.
The optical depth of the inner region from the center to radius $r \leq r_0$ is
\begin{equation}
   \tau_{in} = \tau_0 \left( { r \over r_0 } \right)
             = \tau_0 \left( { v \over v_0 } \right)
\end{equation}
and the optical depth from radius $r_0 = v_0 t$ outwards to radius $r \geq r_0 $ is
\begin{equation}
   \tau_{out} = { \tau_0 \over 7 } \left[ 1- \left( { r_0 \over r } \right)^7 \right]
              =  { \tau_0 \over 7 } \left[ 1- \left( { v_0 \over v } \right)^7 \right].
\end{equation}
If one puts all the material with $v > v_s > v_0$ into a thin shell at radius $v_st$,
the optical depth of the shell is
\begin{equation}
   \tau_{shell} = { \tau_0 \over 6 } \left( { v_0 \over v_s } \right)^7.
\end{equation}
This shows that the detailed arrangement of the outer material has little effect
on the optical depth.
Other changes in the density profile will produce similar, modest
changes in the dimensionless prefactors of these expressions and are unimportant
compared to changes in the opacity, mass, energy, velocity or age of the ejecta. 

The dust opacity is obviously a key quantity, with 
\begin{equation}
   \kappa (\lambda) = { 3 \over 4 \rho_g } { Q(a,\lambda)  \over a }
\end{equation}
where  $\rho_g$ ($\simeq 2.2$ and $3.5$~g~cm$^{-3}$
for graphitic and silicate dusts) is the
bulk density of a grain, $a$ is the grain size, and $Q(a,\lambda)$ is is the usual 
efficiency factor.  Figure~\ref{fig:opacity} shows the visual absorption and scattering
opacities, $\kappa_a$ and $\kappa_s$, for standard graphitic and silicate 
dust as a function of grain size  (\citealt{Draine1984}, \citealt{Laor1993}). 
The total opacity is $\kappa_t = \kappa_a +\kappa_s$
and the effective absorption opacity is
$(\kappa_a \kappa_t)^{1/2}$ due to the increase in path lengths created
by scattering. More or less independent of composition or grain size, the
scale of the V-band opacity is of order $10^4$~cm$^2$/g, as used in the optical
depth scalings given above.  We will treat the absorption as a foreground
screen since the differences between a foreground screen and an unresolved
dust shell are unimportant for our order of magnitude discussion (see
\citealt{Kochanek2012}).

Figure~\ref{fig:expected} shows the V-band optical depth scale $\tau_0$ as a function of
time for a fiducial model with $M_d=0.1 M_\odot$ of dust, an opacity of $\kappa =10^4$~cm$^2$/g,
an explosion energy of $E=10^{51}$~erg, and either $M_e=10M_\odot$ ($v_0=3200$~km/s)
or $1.4M_\odot$ ($v_0=8500$~km/s) to represent ccSNe and Type~Ia SNe, respectively.  
The absolute normalization can be altered by changing the dust mass, 
the opacity, the debris velocity range
over which dust forms or the structure of the self-similar model.  Changes in the
velocity range or the density structure should be relatively unimportant compared to
changes in the dust mass or opacity, which is why we simply use the scale $\tau_0$.
In particular, ccSNe likely produce more dust than $M_d = 0.1 M_\odot$ and Type~Ia
SNe likely produce less.  

The simple $\tau \propto t^{-2}$ power law dictated by expansion will fail at
early times ($t \ltorder 2$~years) when the dust properties are still evolving. It
will also fail at late times as the reverse shock begins to destroy dust (e.g., \citealt{Bianchi2007},
\citealt{Nozawa2007}), 
although this likely only matters on longer time scales than we consider here.  This model also
excludes any dust that might form in the contact discontinuity (a ``cold dense
shell'') between the shocked ejecta and the shocked CSM (e.g., \citealt{Chevalier1994}, \citealt{Pozzo2004}).  

In Figure~\ref{fig:expected},
we have also roughly marked where various sources, many of which we introduced
in \S1,  should lie given our fiducial parameter assumptions.  For ccSNe, we show the  
Type~IIb SNe~2011dh and 1993J, where there are arguments for the detection
of a companion, along with the much older Type~IIb remnant Cas~A.  We also
show SN~1987A, the Crab and G54.1$+$0.3.  For Type~Ia SNe, we show the
very recent SN~2011fe (\citealt{Nugent2011}), along with the SN~1885 (see \citealt{Fesen1989},
\citealt{Fesen2015}), 
Tycho, Kepler, and SN~1006.

The first point to note from Figure~\ref{fig:expected} is that SNR should be optically
thick to dust for their first 
\begin{equation}
    t_{\tau>1} = 55 \kappa_4^{1/2} M_{d,0.1}^{1/2} M_{e10}^{1/2} E_{51}^{-1/2}~~\hbox{years}
   \label{eqn:opthick}
\end{equation}
if the amount of dust formed in anyway corresponds to theoretical expectations.
For the ccSNe,
the theoretically predicted dust mass can be ten times larger.  For the
Type~Ia SNe, the time scale is $t_{\tau>1}\simeq 20 \kappa_4^{1/2}M_{d,0.1}^{1/2}$~years, where the theoretically predicted
dust mass might also be a ten to a hundred times smaller.  Overall, core collapse SNRs should be
optically thick for decades up to a century, and Type~Ia SNRs should be optically
thick for years up to a decade.  

The SN~1987A remnant should still be very optically thick given the
recent Herschel and ALMA estimates that the cold dust mass
is $M_d \simeq 0.5$ to $1.0M_\odot$ (\citealt{Matsuura2011}, \citealt{Indebetouw2014}
\citealt{Matsuura2015}).   \cite{Graves2005}, in their limits on any
surviving binary companion, used much lower optical depth estimates based
on dust masses estimated from the ratio of optical/UV to mid-IR emission 
found by \cite{Bouchet1993} 2172 days after the explosion.
Apparently, much of the dust must have formed at slightly later 
times (see, e.g., \citealt{Wesson2015}). Another important point is that
the bulk of the dust appears to have expansion velocities of  $v \ltorder 1400$~km/s,
putting it well inside the expanding shock between the ejecta and the
surrounding medium.  This might be expected because the outer layers
of the ejecta are the least likely to 
form dust both because they have the lowest metal fractions (i.e. Type~IIP
envelopes) and because they have the highest expansion velocities and so
the most rapidly dropping densities (e.g., \citealt{Kozasa1989}).  
This means that the reverse shock does not start to destroy newly formed
dust until late in the evolution of the SNR.

The best cases for the detection of binary companions to ccSNe are the Type~IIb
SN~1993J and SN~2011dh. 
\cite{Maund2004} report the spectroscopic detection of a early-B
supergiant as the putative binary companion to SN~1993J $9.93$ years
after the explosion, and \cite{Fox2014} argue that new HST observations
taken in late 2011 and early 2014 (so $\sim 19$~years post explosion)
confirm the result.  Both use an estimated extinction of $A_V \simeq 0.6$~mag
derived from photometry of nearby stars in \cite{Maund2004}. 
If we fit the F218W, F275W and F336W magnitudes from \cite{Fox2014} to
the Solar metallicity PARSEC models (\citealt{Bressan2012}), focusing
only on temperature and extinction, we find a slightly smaller estimate
$A_V \simeq 0.3 \pm 0.1$~mag, but there is unmodeled contamination
from SN emission that introduces additional systematic uncertainties.
With a Galactic extinction of roughly $A_V \simeq 0.2$ (\citealt{Schlafly2011}),
we can conservatively use a limit that $A_V \ltorder 0.4$~mag at both
epochs corresponding to $\tau_V \ltorder 0.4$.  For an elapsed time of 
10 or 19 years, we would expect $\tau_V \simeq 30$ and 
$8.2 \kappa_4 M_{d0.1} M_{e10} E_{51}^{-1}$, respectively, 
implying limits on the dust masses of $M_d \ltorder 0.0013$ and 
$0.0049 E_{51} \kappa_4^{-1} M_{e10}^{-1} M_\odot$.  The source also appears to have faded, presumably
due to decaying emission by the SN, without evidence that the 
contribution from the putative companion has increased as it
should if veiled by dust formed in the SN.  

Because SN~2011dh is so much younger, the constraints are correspondingly
tighter. \cite{Folatelli2014} and \cite{Maund2015} analyze the same
HST data taken in August 2014, $3.2$ years after the explosion.  
\cite{Folatelli2014} adopt an extinction of $A_V=0.3$~mag as
found for nearby stars by \cite{Murphy2011} and argue for the
presence of the blue, B star predicted in binary models for the
production of SN~2011dh by \cite{Benvenuto2013}. \cite{Maund2015}
agree with the photometry but argue that some or all of the observed
flux could still be due to the SN.  The expected optical depth
at this epoch is $\sim 290 \kappa_4 M_{d0.1} M_{e10}^{-1} E_{51}^{-1}$,
so a detection of the binary companion with $A_V=0.3$~mag 
implies $M_d \ltorder 10^{-4}E_{51} \kappa_4^{-1} M_{e10}^{-1} M_\odot$. 
This leaves a puzzle as to why some Type~IIb SNe form dust 
(Cas~A), while others apparently do not (SN~1993J and
potentially, SN~2011dh).

\begin{figure}
\centering
\includegraphics[width=0.45\textwidth]{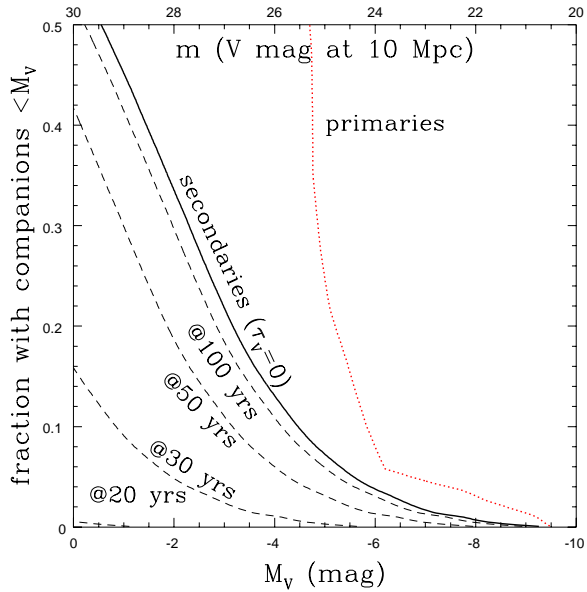}
\caption{ Fraction of all ccSNe with secondaries brighter than $M_V$ assuming no SN dust (heavy black solid,
  $\tau_V=0$). The dashed black lines show the effect of including our standard absorption model
  for times $20$, $30$, $50$ and $100$~years after the SN. The heavy red dotted line shows
  the integral magnitude distribution of the primaries when they explode for comparison.
  This model assumes that all systems were binaries ($F=1$), so the fraction with stellar
  companions at death is $70\%$.  The time scale can be converted to other parameter
  choices following Equation~\protect\ref{eqn:opthick}.  The upper scale shows the corresponding apparent
  magnitudes for a source at a distance of $10$~Mpc.
  }
  \label{fig:magdist1}
\end{figure}

\begin{figure}
\centering
\includegraphics[width=0.45\textwidth]{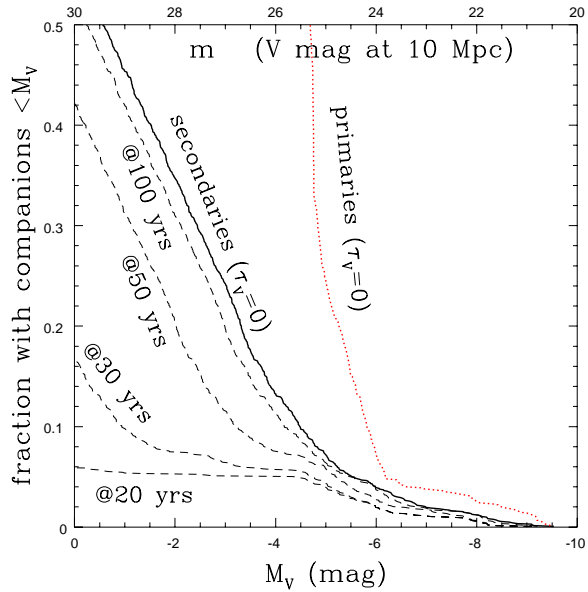}
\caption{ As in Figure~\protect\ref{fig:magdist1} but with no dust formation if the black body
  flux of silicon ionizing photons ($>8.2$~eV) exceeds $Q_* > 10^{49}$~s$^{-1}$,
  roughly matching the companion to SN~1993J.  Many of the most luminous binary
  companions are now unobscured from the start because dust formation is suppressed.
  The companion to SN~1993J has $M_V \simeq -5.9$~mag.
  }
  \label{fig:magdist2}
\end{figure}

\section{The Effects of A Hot Binary Companion on the Debris}

One way to reconcile visible companions in SN~1993J and 
SN~2011dh with dust formation in Cas~A, is that the presence
of a luminous blue companion in SN~1993J and SN~2011dh
suppresses dust formation. 
If the binary companion can photoionize the key elements
for dust formation (Mg, Al, Si, S, Fe and C) sufficiently 
rapidly, then presumably dust formation will be suppressed 
even as the gas becomes cool enough to allow condensation.  

Consider the Type~IIb dust formation model of \cite{Nozawa2010}.
When the star explodes, it is a $4.5 M_\odot$ star which then
ejects $2.94 M_\odot$ of material.  The inner regions of
the ejecta will form grains based on Mg ($0.11M_\odot$, $A=24$, $7.6$~eV), 
Al ($0.01 M_\odot$, $A=27$, $6.0$~eV), Si ($0.11M_\odot$, $A=28$,
$8.2$~eV), S ($0.03M_\odot$, $A=32$, $10.4$~eV)
and Fe ($0.08M_\odot$, $A=56$, $7.9$~eV), while the outer regions will form 
grains based on carbon ($0.11M_\odot$, $A=12$, $11.3$~eV).  
The numbers give the mass of the element in the ejecta, the
atomic mass of the most common isotope, and the first ionization
energy.  ``Mixed'' with the heavier elements is $1.6 M_\odot$
of oxygen ($A=16$, $13.6$~eV), mixed with the carbon is
roughly $1.5M_\odot$ of helium ($24.6$~eV) and there is roughly
$0.08 M_\odot$ of residual hydrogen in the surface layers.
While masses are not reported, nitrogen ($14.5$~eV) and neon 
($21.6$~eV) will also have significant abundances but are
important only to the extent that they absorb UV photons. 
Note that dust formation can be blocked by photoionization
without needing to ionize the most abundant elements 
(H, He, O) or the other common, but ``inert'', elements
(N, Ne) because all these elements have higher ionization
energies than the key dust forming elements.

We can approximate the structure by putting $0.34 M_\odot$ of
silicon in the inner region ($r<r_0$) and $0.11 M_\odot$ of
carbon in the outer region. Using silicon ($A=28$, $8.2$~eV)
roughly matches the number of atoms and the number-weighted 
mean ionization energy of the Mg/Al/Si/S/Fe mixture.  
The first requirement is that the ejecta must be optically
thick when neutral.  For an element with mass 
$M_i = 0.1 M_{i0.1}M_\odot$ in the
ejecta, the photoionization optical depth scale is 
\begin{equation}
  \tau_p = { 15 M_i M_e \sigma_p \over 64 \pi E A m_p t^2 }
     \sim 10^5 { M_{i0.1} M_{e10} \sigma_{17} \over E_{51} A_{28} t_1^2 }
\end{equation}
where $\sigma_p \sim 10^{-17}\sigma_{17}$~cm$^2$ is a
typical photoionization cross section near threshold,
$ A = 28 A_{28}$, and $t = t_1$~years. This means that
in the early phases we can assume that every ionizing
photon is absorbed.  As the photosphere retreats, these 
metal atoms will initially be neutral, so a spectrum
of any binary companion should show a strong absorption
edge near $8$~eV ($1600\AA$).  That the UV spectra of
the companion of SN~1993J from \cite{Fox2014} do not show such an edge
strongly suggests that the Mg, Si, etc.,  have 
all been photoionized.

Absent other sources of ionizing photons, the companion will
form a photoionized region at the center of the SNR.  The
ionizing flux needed to balance recombination interior to $r_0$ is 
\begin{equation}
   Q_0 = { 75 \alpha M_i^2 \over 256 \pi A^2 m_p^2 v_0^3 t^3 } 
     \simeq 2 \times 10^{47} { \alpha_{13} M_{i0.1}^2 M_{e10}^{3/2} \ 
         \over A_{28}^2 E_{51}^{3/2} t_1^3}~\hbox{s}^{-1}
     \label{eqn:need}
\end{equation}
where $\alpha = 10^{-13} \alpha_{13}$~cm$^3$/s is the recombination rate.
This assumes that the elements included in $M_i$ are either
neutral or singly ionized and that any other elements in the inner 
region are little ionized.  This is likely true given the higher
ionization potentials of helium, nitrogen, oxygen and neon.  Photoionizing the
outer region ($r>r_0$) requires $3/13$ of this flux.

A black body produces 
\begin{equation}
   Q_* = { 15 L \over \pi^4 k T } \gamma\left({ h\nu \over k T }\right)
  \quad\hbox{where}\quad \gamma(u) = \int_u^\infty du u^2(\hbox{e}^u-1)^{-1}
     \label{eqn:have}
\end{equation}
photons with energies above $h\nu$.  Since we are interested in the 
photoionization of heavy elements with ionization potentials below that
of hydrogen, the large deviations of stellar spectra from black bodies
beyond $13.6$~eV are not of great importance.  The overall scale
is $15 L /\pi^4 k T = 4.3 \times 10^{49} L_4 T_5^{-1}$~s$^{-1}$ for a 
luminosity $L = 10^5 L_5 L_\odot$ and temperature $T = 10^4 T_4$~K.
For a temperature of $20000$~K, $\gamma(u) = 0.030$, $0.083$, $0.29$,
and $0.66$ for H (13.6~eV), C (11.3~eV), Si (8.2~eV), and Al
(6.0~eV).  Equating Equations~\ref{eqn:need}
and \ref{eqn:have}, the UV photons from the companion can balance
recombination over the full inner region after time 
\begin{equation}
  t_r \simeq 0.34 { \alpha_{13}^{1/3} M_{e10}^{1/2} M_{i0.1}^{2/3} T_4^{1/3}\over
      E_{51}^{1/2} \gamma_{0.1}^{1/3} L_5^{1/3} A_{28}^{2/3}}~\hbox{years}  
\end{equation}
where $\gamma(u) = 0.1 \gamma_{0.1}$.  For comparison, the time scale 
for the companion to produce ionizing photons equal in number
to the number of target atoms in the inner region is
\begin{equation}
    t_p  = { 4 \pi r_0^3 \rho \over Q_* A }
         = { 15 M_i \over 8 A Q_* } 
     \simeq 0.060 { M_{i0.1} T_4 \over A_{28} \gamma_{0.1} L_5 }~\hbox{years}.
\end{equation}
This means that if dust formation occurs on times scales of order
a year, a sufficiently luminous, hot companion can photoionize
the condensible material before dust formation begins.

We can model the growth of the ionized region using the usual simple
model for the growth of a Str\"omgren sphere (\citealt{Osterbrock1989})
modified for the expansion of the medium.  For example,
the growth rate of the photoionized zone in the inner ($r < r_0$) 
region expressed in comoving coordinates $\hat{r}=r/r_0$ is
\begin{equation}
   { d \hat{r} \over dt }
    = { 1 \over \hat{r}^2 t_p } - { \hat{r} \over t_p } \left( { t_r \over t } \right)^3
         - { \hat{r} \over t }
\end{equation}
which has no simple solution.  The first term is the growth of the
photoionized region due to the injection of ionizing photons, the
second term is the losses due to recombination, and the third term
is a Hubble expansion term from working in terms of $\hat{r}$ instead of $r$.
The solution can be extended beyond $r_0$, but the expressions 
become more complicated.  In practice, we considered a model with
a $L=10^5 L_\odot$ and $T=20000$~K black body source ionizing an
inner region containing $0.34 M_\odot$ of silicon 
($A=28$, $\gamma=0.29$) to be ionized, and an outer region
containing $0.11 M_\odot$ of carbon ($A=12$, $\gamma=0.083$).
Using silicon is a reasonable proxy for the actual mixture of 
Mg/Al/Si/S/Fe.  We assume that the silicon and carbon are 
either neutral or singly ionized and that all other elements
with their higher ionization potentials are nearly neutral.  
In this model, the remnant
is photoionized in 530 days starting from neutral gas.  The
inner region is photoionized in 200 days, while the photoionization
of the outer region is slowed because of the higher ionization
potential of carbon and the overestimation of the recombination
rate created by putting all the carbon in the outer region. 
With one-third the luminosity, it takes four years and with
three times the luminosity it takes 240 days.

The real evolution would be more complex since we should
follow the ionization of all the species. We have also
neglected the emission by the SN itself, which includes
a locally ionizing component because of the radioactive
decays, the evolution of the photosphere, collisional
ionization, and any ionizing radiation from the 
expanding shocks.   Most of these effects should accelerate
the photoionization of the ejecta by the secondary 
because it means the ejecta are either starting from
a partially ionized state or there are additional
sources of ionization.  Simply starting the process later
(e.g., at the end of the plateau phase) has almost
no effect on the time at which photoionization is 
complete because the faster initial
growth due to the lower densities compensates for the
lost time at the start.  For example, solutions to our simple photoionization
model started 90 days after the start of the expansion
simply merge onto the previous solution.

We chose these parameters because they roughly match the estimated 
properties of $T \simeq 10^{4.3\pm 0.1}$~K 
and $L \simeq 10^{5.0\pm0.3}L_\odot$ for the
companion to SN~1993J(\citealt{Maund2004}).
\cite{Fox2014} agree with this temperature estimate but provide
no independent estimate of the luminosity.  \cite{Folatelli2014}
do not directly estimate a temperature and luminosity for their
proposed companion to SN~2011dh but compare to the models of
\cite{Benvenuto2013}.  In these models the secondary ranges
from $T \simeq 22000$~K and $L = 10^{3.8}L_\odot$ if the 
accretion efficiency is low to $T_* \simeq 39000$~K and
$L= 10^{4.8}L_\odot$ for high accretion efficiencies.  
\cite{Folatelli2014} favor the lower efficiency solutions
with the lower luminosity and temperature companion.
A star with the estimated proprieties of the companion to
SN~1993J or those of the hotter more luminous models for SN~2011dh
would photoionize the most important dust producing species
on the time scales that \cite{Nozawa2010} find for dust formation
in Cas~A (300-350, 350-500 and 600-700~days for
the carbon, silicate/oxide, Si/FeS grains, respectively).
The cooler, lower luminosity models for SN~2011dh probably
could not do so.  The SN~1993J remnant would also be
photoionized well before \cite{Fox2014} obtained their UV HST 
spectra in 2012, which would explain why the are no signs of 
strong absorption bluewards of $\sim 1600\AA$.  

\section{An Illustrative Model}

In \cite{Kochanek2009} we considered a simple model for the expected properties of 
binary companions to ccSNe ignoring binary interactions.
Here we take one of these cases to illustrate the
consequences of the evolving dust distribution.  We assume that the distribution
of primary masses $8M_\odot < M_p < 100 M_\odot$ is Salpeter, $dN/dM_p \propto M_p^{-x}$ with $x=2.35$
and that fraction $F$ of the primaries have binary companions distributed in
mass $M_s$ as $f(q)$ with $q_{min} < q = M_s/M_p < q_{max}$ where 
$\int dq f(q) \equiv 1$.  The joint distribution of the primary and 
secondary stars in mass is
\begin{equation}
   { d N \over dM_p dM_s } = F M_p^{-x-1} f(q)
\end{equation}
with a secondary mass distribution of
\begin{equation}
   { d N \over dM_s } = F f_q M_s^{-x}
   \quad f_q = \int_{q_{min}}^{q_{max}} q^{x-1} f(q) dq.
\end{equation} 
An observed ccSN can be the collapse of a single star that was
never in a binary ($f_{single}= (1-F)/(1+F f_q)$), the
primary of a binary ($f_p = F/(1+F f_q)$), or the secondary
of a binary ($f_s= F f_q/(1+F f_q)$).  The fraction of ccSNe
with a stellar binary companion is just $f_p$, since when the
secondary of a binary explodes the primary is now a compact
object (and the binary may have been disrupted).  

Here we will
just consider the case with $f(q)$ constant over $0.02 < q < 1$.
For this uniform model, $f_q = 0.434$ and 70\% (41\%) of ccSNe
have stellar binary companions if $F=1$ ($F=1/2$). We show
results for $F=1$ as they can be trivially modified for the
addition of stars which are not in binaries.
We used the Solar metallicity v1.2S PARSEC 
(\citealt{Bressan2012}, \citealt{Tang2014}) isochrone models  
to provide estimates of the V band magnitudes for stars of
a given age and mass.  For these models, the primaries 
become steadily more (V band) luminous up to $ \sim 30 M_\odot$
and then rapidly become fainter as mass loss leads to higher
temperatures and larger bolometric corrections.    

Figure~\ref{fig:magdist1} shows the results for all SN
with binary companions.  Most secondaries are fainter,
blue, main sequence stars simply as a consequence of
their lower masses, as already noted in \cite{Kochanek2009}.
For this simple model, only $\sim 10\%$ of SNe
have observable ($V \gtorder 26$~mag at 10~Mpc) secondaries
even though most SNe (70\% for the $F=1$ binary fraction used
here) occur in systems with a stellar companion.  With
the addition of dust, companions should almost never be
observable until decades after the SN.  However, this
leads to the interesting observational possibility of
searching for slowly reappearing stars at the sites of 
decades old SNe. 

Now, suppose that dust formation is suppressed whenever 
$Q_* > 10^{49}$~s$^{-1}$ (Equation~\ref{eqn:have}).
This is roughly the flux of silicon ionizing photons
from the companion of SN~1993J.  As shown in Figure~\ref{fig:magdist2},
many of the more luminous companions are now
easily observed because dust formation has been
suppressed.  For a threshold of $Q_* > 10^{49}$~s$^{-1}$,
dust formation is suppressed in roughly 8\% of the
binary systems.  If the threshold is even modestly
higher, the mechanism does not work, since the fraction
drops to 3\% for a threshold of $Q_* > 10^{49.5}$~s$^{-1}$ and
is negligible for $Q_* > 10^{50}$~s$^{-1}$.  Lowering
the threshold rapidly raises the fraction to 15\% for $Q_* > 10^{48.5}$~s$^{-1}$ 
and 25\% for $Q_* > 10^{48}$~s$^{-1}$.  
Not all of the visually luminous stars
are unobscured, since there are some companions that
are luminous but cool. 
The companion to SN~1993J should have $M_V \simeq -5.9$~mag,
and it lies in the magnitude range of Figure~\ref{fig:magdist2}
where the companions can affect dust formation (as expected).

\section{Discussion}

If SNe form dust as predicted in theoretical models and observed in nearby
SNRs, then the binary companions to SNe, particularly ccSNe, should
initially be heavily obscured.  The decreasing optical depth
depth due to expansion should lead to a steady brightening that
should be a nearly unique signature of a binary companion.
The inferred optical depths then
provide an estimate of the amount of newly formed dust that is 
independent of the dust temperature.  Absorption by dust should 
be observable (optical depths $\gtorder 0.01$) for decades to
roughly a century after the explosion.  
The local SNR that might be probed through dust absorption
are SN~1987A, Cas~A and the Type~Ia SN~1885 in M31.  

The present day optical
depth scale for SN~1987A is $\tau_0 \sim 3 \kappa_4^{-1} M_{d0.1}$ and
current estimates are that $M_d \simeq 0.5$-$1.0M_\odot$
(\citealt{Matsuura2011}, \citealt{Indebetouw2014},
\citealt{Matsuura2015}).  Even for these high dust
masses, SN~1987A is approaching the point where it 
is feasible to search for either a surviving companion 
or a fortuitously obscured background star in the
near-IR. Many models for SN~1987A invoke binary
interactions, although frequently with a final
merger, to explain either the explosion of a 
blue supergiant or the structure of the surrounding winds 
(e.g., \citealt{Podsiadlowski1989}, \citealt{Podsiadlowski1992}, 
\citealt{Blondin1993}, \citealt{Morris2009}). If there
is a surviving companion, the observed mid-IR dust 
luminosity limits its luminosity to be $\ltorder 10^2 L_\odot$.

While a dwarf (single degenerate) companion to SN~1885A is
probably not observable even if present, 
the remnant of SN~1885A is observed as a spatially
extended metal line absorption feature against the stars 
in the bulge of M~31 (\citealt{Fesen1989}, \citealt{Fesen2015}).
These authors comment that there is no obvious continuum 
absorption by the SNR, implying little dust, but do not present a
quantitative limit.  The present
day optical depth scale for SN~1885A of 
$\tau_0 \sim 0.02 \kappa_4^{-1} M_{d0.1}$ is consistent
with this comment and we make a quantitative estimate
in Auchettl \& Kochanek (2017).  

The present day optical depth scale for Cas~A should be similar
to that of SN~1885, with 
$\tau_0 \sim 0.03 \kappa_4^{-1} M_{d0.1}$, because the
slower expansion of a ccSN compensates for Cas~A's greater 
age. With dust mass estimates approaching $M_d \sim M_\odot$
(\citealt{Barlow2010}, \citealt{deLooze2016}), the optical
depth of the SNR could be significant.  This suggests
trying the method of \cite{Trimble1977}, comparing the
colors of stars superposed on the SNR to those of other
nearby stars.  This method is probably infeasible because
of the limited numbers of stars and because even modest
spatial inhomogeneities in the $A_V \sim 5$-$6$~mag of 
extinction towards Cas~A (\citealt{Hurford1996}) will
confuse the measurements.  With some patience, however,
it is feasible to measure the optical depth through
the time variability of the extinction of background
stars.  Over a decade, the optical depth should drop
by $\sim  0.002 \kappa_4^{-1} M_{d0.1}$ which is well
within the capabilities of difference imaging methods
if $M_d \simeq M_\odot$.  This depends on the existence 
of a (non-variable) background star since Cas~A seems to not have had
a binary companion (\citealt{Kochanek2017}), but is
an otherwise straight forward experiment.

The best cases for the detections of binary companions are 
the B stars proposed for the Type~IIb's SN~1993J (\citealt{Maund2004}, 
\citealt{Fox2014}) and SN~2011dh (\citealt{Folatelli2014},
but see \citealt{Maund2015}). Photometry of both progenitors
showed blue excesses (e.g., \citealt{Aldering1994}, \citealt{Arcavi2011},
\citealt{Bersten2012}) and most models for Type~IIb
SNe invoke binary mass transfer with predicted secondary
properties similar to those of the observed candidates 
(e.g., \citealt{Podsiadlowski1993},
\citealt{Stancliffe2009}, \citealt{Claeys2011}, \citealt{Benvenuto2013}).
However, the colors of the candidate stars require very little extinction,
and so imply very stringent limits on the dust mass of $M_d \ltorder 10^{-3}$
and $10^{-4}  \kappa_4^{-1} M_{e10}^{-1} M_\odot$, respectively.  Theoretical
models by \cite{Nozawa2010} predict the formation of $M_d \simeq 0.1M_\odot$
of dust, and \cite{Bevan2016} infer the presence of $M_d \simeq 0.1$-$0.3M_\odot$
of dust in SN~1993J in order to model the asymmetries of optical emission lines. 

Either these companion identifications are incorrect, or these Type~IIb
SNe formed negligible amounts of dust.  Inhomogeneities can lead to particular lines
of sight having more or less extinction, but this should be a matter
of degree rather than any light of sight being genuinely transparent.
It is also implausible as an argument for rendering
both sources visible. Cas~A was also a Type~IIb SNe (\citealt{Krause2008},
\citealt{Rest2008}) and it both lacks a companion (\citealt{Kochanek2017})
and contains a significant amount of dust ($M_d \sim 0.1$-$0.8M_\odot$,
\citealt{Barlow2010}, \citealt{deLooze2016}).  The companion stars
invoked for SN~1993J or SN~2011dh would be $R \sim 16$~mag in Cas A
and impossible to miss (see \citealt{Kochanek2017}).

This circle can be squared by the fact that a luminous, hot, stellar
companion can suppress dust formation by photoionizing the ejecta
sufficiently rapidly.  A least in our simple model of the process,
the proposed companion to SN~1993J can do so, and this may also be
true of the proposed companion to SN~2011dh.  
Detailed calculations
of the full multi-element, time-dependent photoionization process
are required to determine the exact threshold for suppressing
dust formation. In addition to
photoionization, the abundant soft UV photons from
a hot companion further inhibit dust formation because they
easily destroy very small grains by stochastically heating
them much to higher temperatures than predicted by the 
equilibrium temperature (\citealt{Kochanek2011}, \citealt{Kochanek2014}).  
For SN~1993J in particular, there are also $\sim 10^{39}$~ergs/s of
X-rays being generated by the SN shock expanding into the 
circumstellar medium at the time when dust would form.  Some
of these X-rays must also contribute to reionizing the interior.

If the secondary has reionized the ejecta, a late time
spectrum should show a peculiar, low ionization, heavy
element emission line spectrum.  At present,
spectra of SN~1993J are dominated by emission lines from higher
ionization states and with a ``boxy'' shape that indicates
they are largely produced in a shell associated with the
shock region (e.g., \citealt{Fransson2005}).  The companion
is too cool to contribute to producing these highly ionized
states.  Where a spherical
shell produces exactly a top hat velocity distribution,
with $dP/dv$ constant out to the expansion velocity of the shell,
recombination lines produced by the companion fully
photoionizing the interior ($r < r_0)$ would be more
``paraboloidal'' with a line-of-site velocity distribution
$dP/dv \propto 1-v^2/v_0^2$.
Adding the contribution from the exterior
($r>r_0$) would add fainter, broader wings. This
contribution to the spectrum of SN~1993J could become
more visible as the contribution from circumstellar
emission slowly fades.

Since binary companions to SNe should be common and will frequently 
be hot, main sequence O/B stars (e.g., \citealt{Kochanek2009}), variability 
in the binary properties of SNe might help to explain the some of 
the seeming incoherence of the evidence for dust formation in SNe.  
In our simple ``passively evolving'' model, dust formation can be 
suppressed in 5-25\% of ccSNe that are stellar binaries at death.
But the effects of the companion are likely more complex. Since
the dust forming elements are stratified in radius and have a
range of different ionization potentials, less luminous binary
companions will still modify which dusts form even if dust 
formation is not completely suppressed.  For example, silicate
dust formation is easier to suppress than carbonaceous dust
formation because silicon is both closer to the center of the
SNR and more easily ionized than carbon.

If many ccSNe form dust, and this still seems inevitable given
the physics of dust formation and the observed dust masses in
nearby SNR, then almost all limits on the existence of binary
companions from post-explosion imaging are invalid unless the
companion is sufficiently soft-UV luminous to suppress dust 
formation.  Binary companions must generally be identified by
the appearance of a star at the location of the SN decades 
after the explosion.  

Dust formation can also have interesting consequences for
interpreting the constraints on the progenitors of the
stripped Type~Ib/c ccSNe.  Because of bolometric corrections,
a secondary star to these SNe can be more optically
luminous than the primary (see \citealt{Kochanek2009}).
This means that in pre-explosion images, the observed
source can be the secondary rather than the primary.
Without dust formation, such a mistaken identification
is easily rectified because the secondary will still
be present in post-explosion images.  However, if dust
forms, the secondary will also be invisible in the
post-explosion images for a long period of time and this
may be incorrectly interpreted as confirming that the 
pre-explosion source was the primary.

This is presently of greatest relevance 
to the best candidate for a Type~Ib progenitor, iPTF13bvn (\citealt{Cao2013},
\citealt{Groh2013}, \citealt{Bersten2014}, \citealt{Fremling2014},
\citealt{Eldridge2015}, \citealt{Eldridge2016},
\citealt{Folatelli2016}).   Present
models for iPTF13bvn prefer scenarios in which the primary
was the more visually luminous star, but also predict that
the secondary is likely observable as the 
SN fades (e.g. \citealt{Eldridge2016}).
Like most other ccSNe, Type~Ib SN are predicted to create        
significant amounts of dust (\citealt{Nozawa2008}).  For
the predicted properties of the secondary in the existing models 
($T \simeq 20000$~K and $L < 10^{4.3}L_\odot$), dust
formation is unlikely to be suppressed.  Thus, if no 
secondary is detected, nothing can be said about its
properties because it may simply be heavily obscured.  
If a secondary is detected, than iPTF13bvn
becomes another example of an SN making negligible 
amounts of dust ($M_d \ltorder 10^{-3}M_\odot$) and the 
companion should be at the maximum possible luminosities
predicted by \cite{Eldridge2016}.

\section*{Acknowledgments}

CSK thanks J.J. Eldridge, R. Pogge and T. Thompson
for discussions.
CSK is supported by NSF grants AST-1515876 and
AST-1515927.

\end{document}